\documentclass[prl,twocolumn]{revtex4}
\usepackage{epsfig}
\usepackage{latexsym,amssymb,graphicx,caption}

\newcommand{\be}{\begin{equation}}
\newcommand{\ee}{\end{equation}}
\newcommand{\bea}{\begin{eqnarray}}
\newcommand{\eea}{\end{eqnarray}}

\begin{document}

\title{Impulse distributions in dense granular flows: signatures of large-scale spatial structures}

\author{Allison Ferguson}
\author{Ben Fisher}
\author{Bulbul Chakraborty}

\affiliation{Martin Fisher School of Physics, Brandeis University,
Mailstop 057, Waltham, Massachusetts 02454-9110, USA}

\begin{abstract}
In this paper we report the results of simulations of a 2D gravity
driven, dissipative granular flow through a hopper system.
Measurements of impulse distributions $P(I)$ on the simulated
system show flow-velocity-invariant behavior of the distribution
for impulses larger than the average impulse $<I>$. For small
impulses, however, $P(I)$ decreases significantly with flow
velocity, a phenomenon which can be attributed exclusively to
collisions between grains undergoing frequent collisions.
Visualizations of the system also show that these frequently
colliding particles tend to form increasingly large linear
clusters as the flow velocity decreases. A model is proposed for
the form of $P(I)$, given distributions of cluster size and
velocity, which accurately predicts the observed form of the
distribution. Thus the impulse distribution provides some insight
into the formation and properties of these ``dynamic'' force
chains.
\end{abstract}


\maketitle
\paragraph{Introduction}Granular materials exhibit a wide spectrum
of behavior ranging from gaseous to liquid to solid.  Remarkably,
all of these phases of granular matter respond to external stimuli
in a manner strikingly different from ordinary fluids and solids
\cite{jaeger96,kadanoff99}.
In static granular piles, the spatially inhomogeneous manner in
which stress is transmitted from the bulk of the pile to the
boundary has been well documented\cite{liu95,howell99}.
Experiments have shown that the force distribution $P(f)$ at the
walls is exponential at large forces and exhibits a plateau at
small forces\cite{mueth98}. In addition, the highly stressed
grains in static packings are organized into linear structures
termed ``force chains''\cite{liu95,howell99}.  The appearance of
similar large scale structures in flowing granular matter would
have significant implications for both continuum theories and
descriptions of jamming in non-thermal systems.  A recent proposal
for a unified picture of jamming in thermal and non-thermal
systems suggests that jamming occurs due the formation of force
chains whose presence is signalled by the appearance of a plateau
in $P(f)$ \cite{ohern01}.  A continuum description of steady-state
flow on an inclined plane models the system as a collection of
transient 1D chains immersed in a viscous fluid\cite{mills99}.
Indeed, transient ``clusters'' have been identified experimentally
in granular surface flows \cite{bonamy02} and shear
flows\cite{miller96}, and simulations of chute flows have shown
evidence for a plateau in $P(f)$ as the system approaches
jamming\cite{silbert02}.

Recent experiments have been performed in a two dimensional hopper
geometry to  explore the presence of incipient force chains in
purely collisional gravity-driven flow \cite{longhi02}.
Measurements of the time trace of the impulse delivered to a
transducer placed at the side wall of the hopper have shown that
the distribution of impulses, $P(I)$, displays an exponential
decay at large $I$, as for the case of static materials.  This
exponential form of the distribution is maintained for all flow
velocities from the largest measured ($v_f$ = 60.0 cm/s) to the
minimum flow velocity prior to the point at which the system no
longer exhibits sustained flow ($v_f$ = 9.4 cm/s). However, at
small $I$, $P(I)$ develops an upward trend which becomes
increasingly more evident as the flow velocity decreases.

In this letter, we report the results of event-driven simulations
of a system of inelastic, monodispersed hard disks in the
experimental geometry of Ref.~\cite{longhi02}.  Our simulations
provide clear evidence of an increasing proportion of collisions
with {\it small impulses} as the flow velocity is decreased.  This
results in the formation of a plateau in $P(I)$ as the minimum
flow velocity for sustained flow is approached.  In addition, we
observe the formation of clusters of disks which collide
``frequently'' and are reminiscent of the ``collapse strings''
observed in freely cooling granular matter\cite{mcnamara96}. We
present a model calculation which strongly suggests that the
increase of small impulse events is associated with the growth of
these clusters.

\paragraph{Simulations}The grain dynamics used in the simulations are
as in Ref.~\cite{denniston99} (momentum is conserved and at each
interparticle collision the energy loss is proportional to $(1-\mu^2)$
where $\mu$ is the coefficient of restitution). To ensure that the
pressure is independent of the height the side walls must absorb some
vertical momentum, therefore we impose the condition that collisions
with the walls are inelastic in the tangential direction.  The flow
velocity is controlled similarly to the experiments, by adjusting the
width of the hopper opening.  However, as we wish to observe the
system over many events, particles exiting the system at the bottom
must be replaced at the top to create sustained flow.  This necessitates
the introduction of a probability of reflection $p$
at the bottom of the hopper (this would be equivalent to the presence
of a sieve in experiment).  In our case, it provides another parameter
with which we can tune the flow velocity. Typically, our simulations
were done on systems of 500 disks, with $\mu$ = 0.9 and $p$ = 0.4.
The simulation was run for $2\times 10^6$ events for each flow
velocity, with $1.5\times10^5$ discarded initially to allow the system
time to reach steady state.

\begin {figure}
  \centering
  \includegraphics [origin = c, scale = 0.34,  angle = 270]{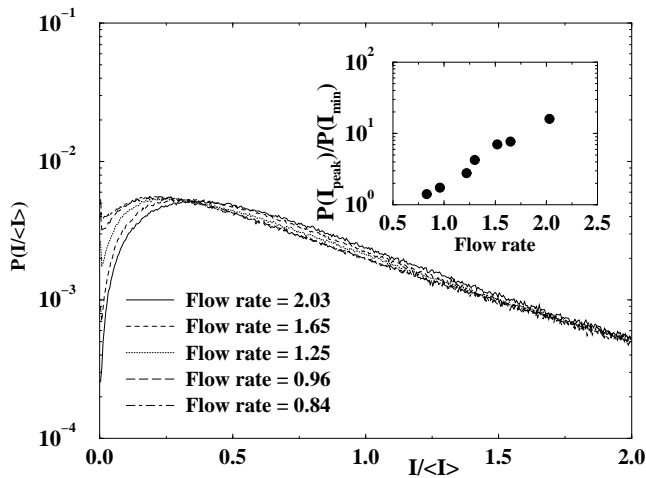}
  \caption{Impulse distribution scaled to the average impulse, ${<}I{>}$, with flow velocity decreasing from the bottom curve to the top curve. The inset shows the ratio of $P(I_{peak})$ to $P(I_{min})$  as a function of the flow velocity.}
  \label{impdist}
\end{figure}

\paragraph{Simulation Results} The physical quantity that is most closely
related to $P(f)$ in our flowing hard-disk system is the
distribution of impulses transferred at each collision, $P(I)$.
Unlike the experiments done in  Ref.~\cite{longhi02}, where the
measurements were made only at the wall, we measure the magnitude
of the momentum transfer at each collision event (note that the
impulse and flow velocity are measured in units such that disk
diameter $d$, disk mass $m$ and acceleration due to gravity $g$
are all equal to one). We observe an exponential form of $P(I)$ at
impulses larger than the average impulse (Fig.~\ref{impdist}) for
the full range of flow rates. After scaling the impulse by the
average impulse, $<I>$, the curves collapse onto each other for
large impulses.  The small impulse behavior of the distribution,
however, changes markedly with changes in the flow velocity. As
the flow velocity decreases, the height of the distribution at a
minimum impulse ($I_{min} = 0.0075<I>$ for this distribution)
increases and $P(I)$ begins to develop a plateau at small $I$.  If
the ratio between $P(I)$ at the peak impulse to $P(I)$ at the
minimum impulse $I_{min}$ is calculated and plotted against the
flow velocity, a dramatic  increase is observed
(see inset to Fig.~\ref{impdist}).  Extrapolating to small flow velocities
indicates that this ratio becomes one at a non-zero flow rate.

It should also be noted that the height of the distribution at $I = 0$
makes a sharp upturn which becomes more pronounced as the flow
velocity decreases (the upturn is not present for the fastest flow
velocity).  While we do not fully understand the reasons for this
phenomenon, we believe it is not the same upward trend visible in the
experiments.  That feature was a smooth continuation of the large
impulse curve, while in our distribution a peak is seen.  In addition,
our upturn occurs only at $I = 0$, regardless of the bin width used in
constructing the distribution.  Such small impulses are not likely to
be distinguishable in experiment due to the resolution limit imposed
by the transducer. For the remainder of our analysis, we will consider
the impulse distribution without the $I = 0$ point.

The basic form of the impulse distribution, a peak at a finite
value of $I$ and an exponential tail can be understood on the
basis of a flow of uncorrelated hard disks.  If the disks were
uncorrelated, the impulse distribution would be a convolution of
the individual momentum (velocity) distributions.  Since there is
an average flow velocity, this would give rise to a peak in the
distribution.  The exponential tail would be observed if the
individual velocity distributions also had  exponential tails. The
velocity distribution observed in our simulations can, to a first
approximation, be described by Gaussians with exponential tails.
This argument would then suggest that the exponential tail arises
from uncorrelated particles and is a consequence of the shape of
the velocity distribution.  Viewed from this perspective, the lack
of change in shape of $P(I)$ at large $I$ as the flow approaches
jamming is not surprising. We will discuss this further in the
context of a simple model to be presented below.

The change in $P(I)$ at small $I$ is reminiscent of the universal
jamming scenario\cite{ohern01}  although the direction of the
change (filling up at small impulses) is different from the one
observed in Lennard Jones systems and foams where the probability
of finding small forces decreases as the system nears the jamming
transition. Nevertheless, the idea that changes in the
distribution at small impulses or forces could be related to the
appearance of structures akin to force chains is intriguing.

\begin {figure}
  \centering
  \includegraphics [scale = 0.36]{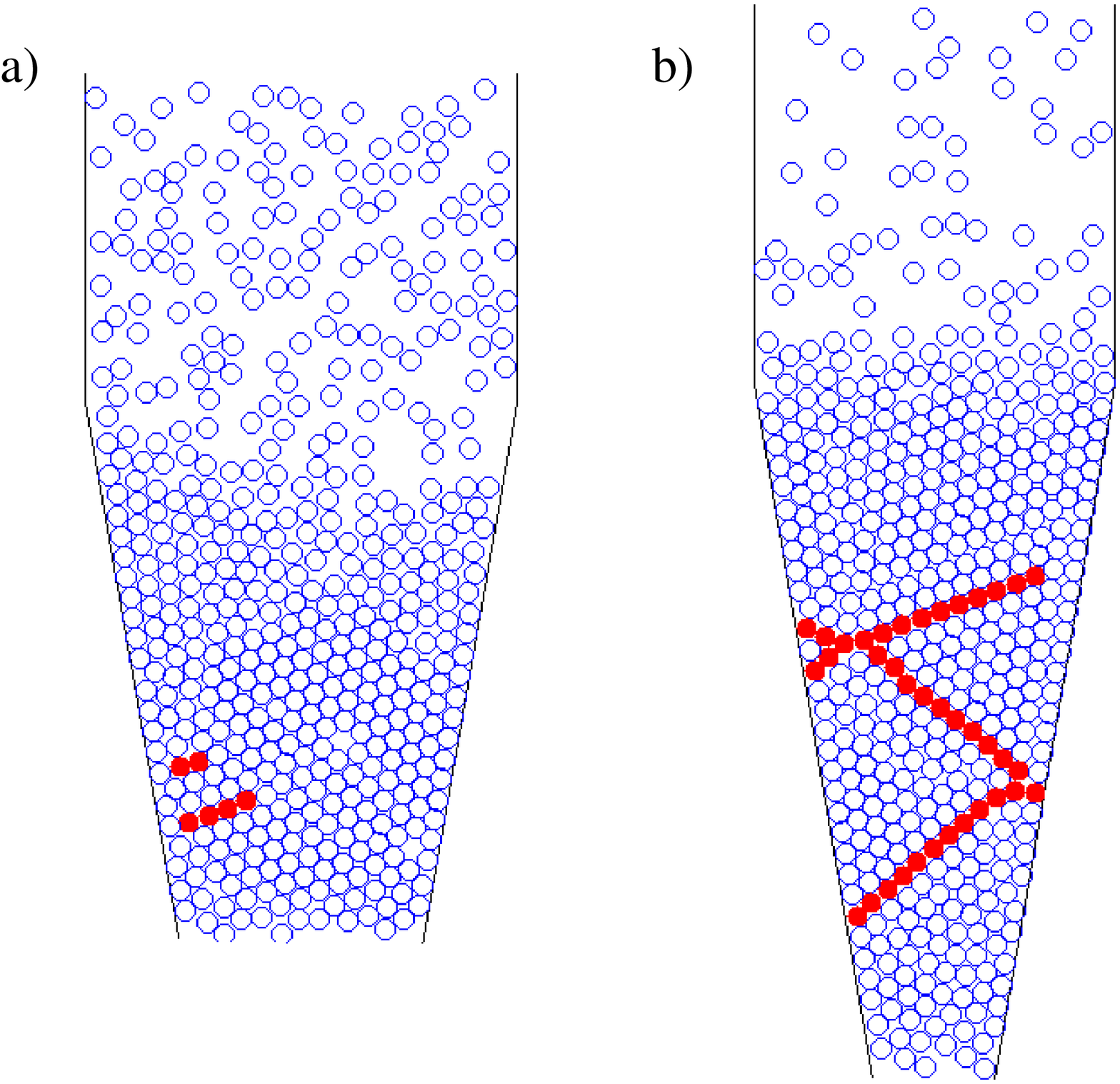}
  \includegraphics [origin = c, scale = 0.34, angle = 270]{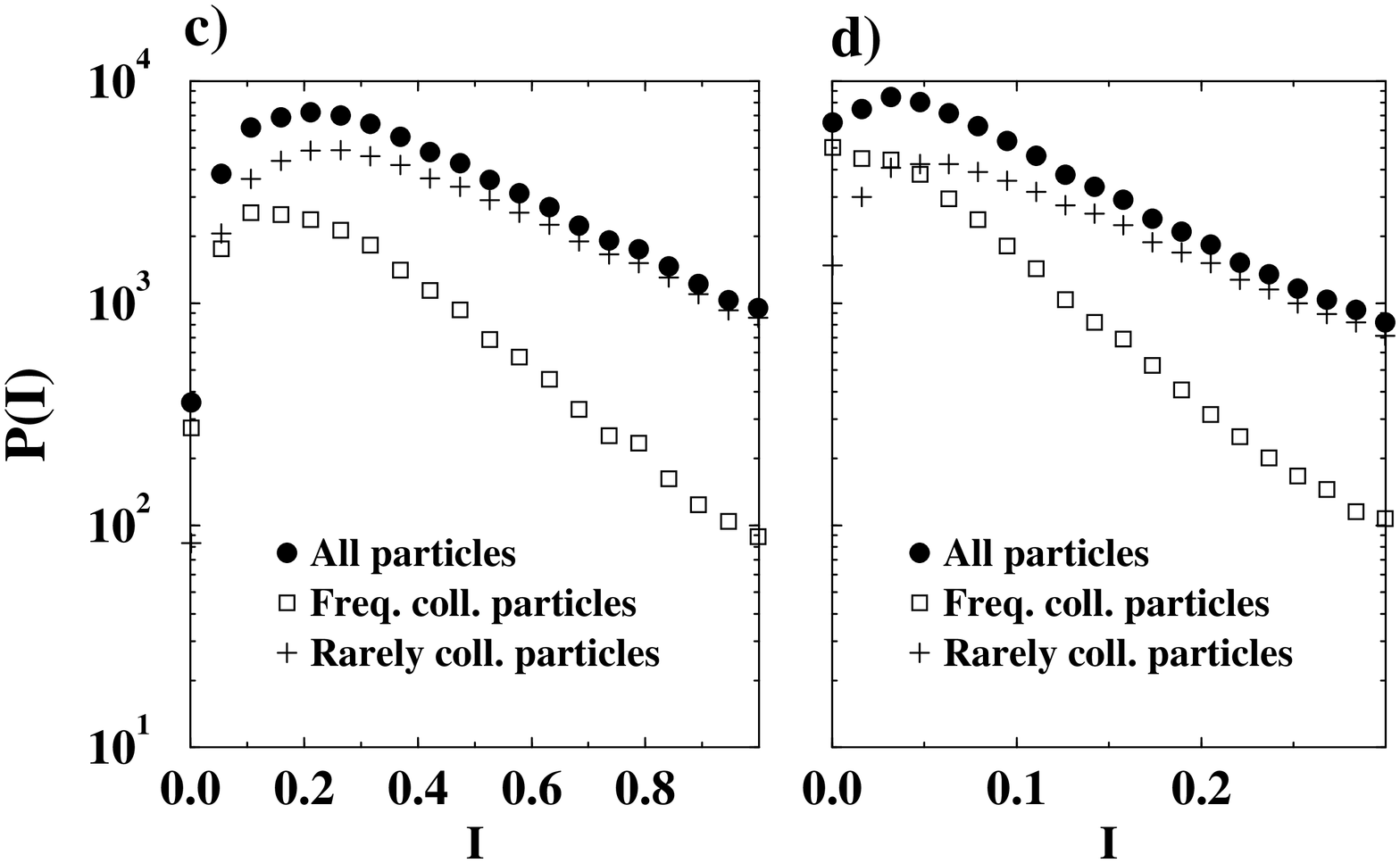}
  \caption{(a) (b) Sample image of simulation for $v_f$ = 2.03 (a) and
  for $v_f$ = 0.96 (b).  (c) (d) Impulse distributions for all
  particles, frequently colliding particles and rarely colliding
  particles for $v_f$ = 2.03 (c) and for $v_f$ = 0.96 (d). Note that
  these distributions are unnormalized. } \label{picts}
\end{figure}

To explore a possible connection between the changes in $P(I)$ and
the appearance of spatial inhomogeneities, we considered a question first
asked in studies of inelastic collapse in freely cooling granular
gases\cite{mcnamara94}, {\it ''How many collisions does a given grain undergo in a fixed number of events?''}  We define a minimum frequency of
collision for a given particle as $\omega_{0}=\frac{1}{50\overline{\tau}}$
(where $\overline{\tau}$ is the average time between events for a given flow
velocity) and then identify those particles undergoing collisions with a frequency
$>\omega_{0}$ as {\it frequently colliding}. This allows us to define a minimum
number of collisions a frequently colliding particle must undergo in a given number
of events.  We then construct an image of our system at every 1000 events, and
color all disks satisfying the criteria (see Fig.~\ref{picts}a,b).  As we decrease
the flow velocity, the frequently colliding particles form increasingly larger
linear clusters (compare Fig.~\ref{picts}a, where $v_f$ = 2.03 in our units
or 35.6 cm/s and Fig.~\ref{picts}b, where $v_f$ = 0.96 or 16.9 cm/s). These
1D structures observed in our simulation are reminiscent of the transient solid chains
postulated by the hydrodynamic model of Ref.\cite{mills99}.
Comparisons of the impulse distribution of the frequently
colliding particles and the impulse distribution of the remaining
``rarely colliding'' particles  (Fig.~\ref{picts} c,d) reveals that
it is the contribution from the frequently colliding particles
which causes the height of the total impulse distribution at
$I_{min}$ to increase in the manner seen in Fig.~\ref{impdist}a.
The large impulse behavior appears to be dominated by the rarely
colliding particles. These observations are relatively insensitive
to changes in $\omega_{0}$ provided it is larger than a minimum
value.

\begin {figure}
  \centering
  \includegraphics[scale = 0.2]{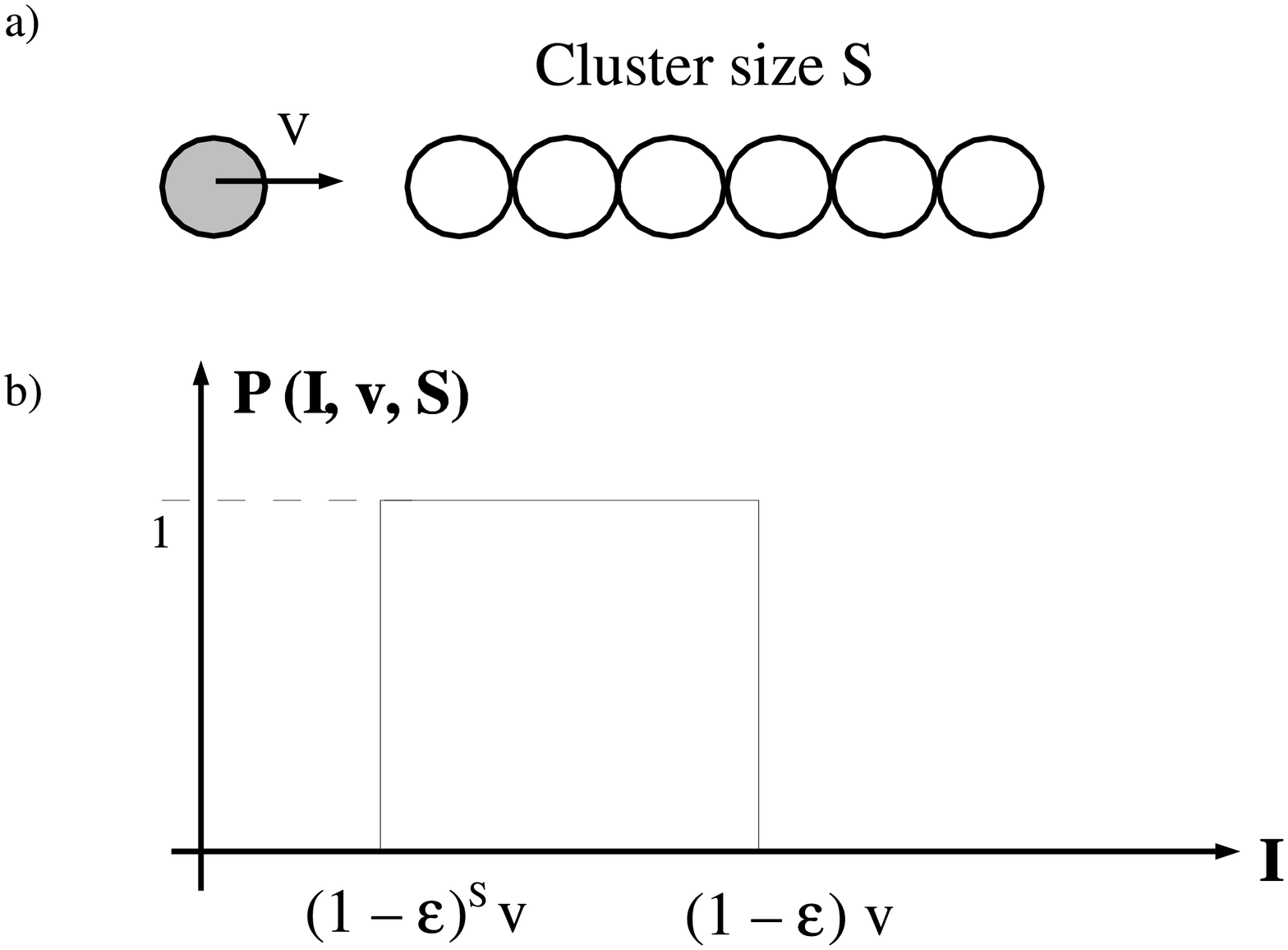}
  \caption {(a) Particle of speed $v$ incident on 1D cluster of $S$ stationary particles. (b) Resulting form of $P(I,v,S)$.}
  \label{chain}
\end{figure}

\paragraph{Impulse Distribution Model}
A possible connection between the shape of the impulse distribution
and the development of linear clusters of frequently colliding particles
can be drawn by investigating the following model.  Consider a one
dimensonal cluster of particles which are all moving with the same
velocity.  Now if another particle, travelling at some speed $v$
relative to the cluster (see Fig.~\ref{chain}a) collides inelastically
with one end of the chain, the impulse associated with that collision
will be $(1-\epsilon)v$ where $\epsilon$ is related to the coefficient
of restitution by $\epsilon = \frac{(1 - \mu)}{2}$\cite{du95}. If the chain was
only comprised of one grain, the impulse distribution $P(I,v,1)$
(given an initial incoming speed $v$) would be a single spike of
height 1 located at $(1-\epsilon)v$.  For two particles, $P(I,v,2)$
 would be a bar of height 1 with limits at $(1-\epsilon)^2v$
and $(1-\epsilon)v$.  Continuing this argument for clusters containing
$S$ particles, then as $S$ becomes very large, the leftmost limit of $P(I,v,S)$
 approaches zero (Fig.~\ref{chain}b).

Given a distribution of speeds $P(v)$ for the incident particle,
and a distribution of cluster sizes $P(S)$, the total impulse
distribution $P(I)$ is:
\begin{equation}
P(I) = \int dS dv P(v) P(S) P(I,v,S)
\label{impulse}
\end{equation}

If the cluster size distribution falls off sharply, then the shape of $P(I)$ will
essentially reflect the shape of $P(v)$.  If, however, the cluster size
distribution becomes broad, then the small impulse end of $P(I)$ will flatten out
reflecting the nature of $P(I,v,S)$ for large $S$.

\begin {figure}
  \centering
  \includegraphics [scale = 0.28]{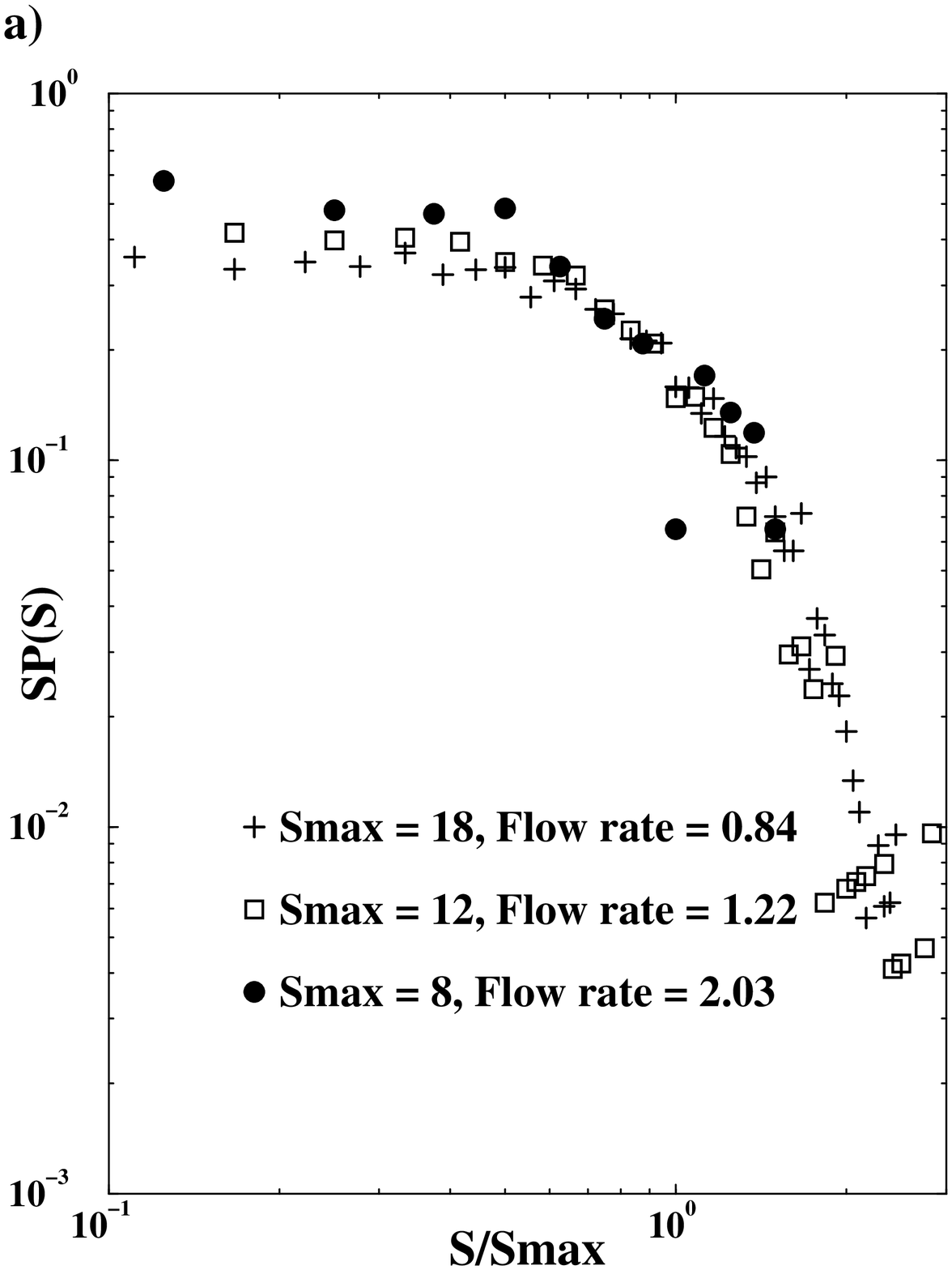}
  \includegraphics[scale = 0.3]{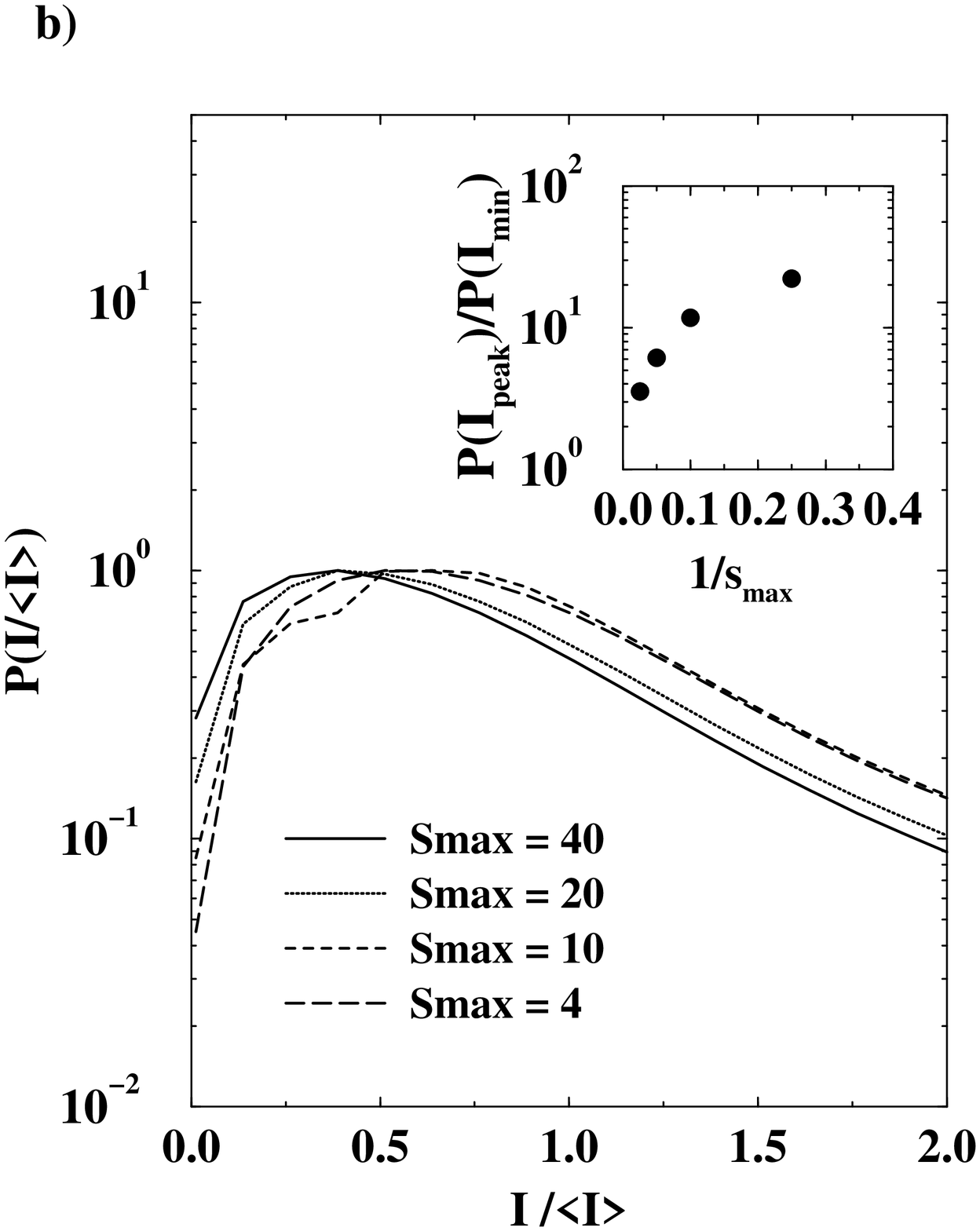}
  \caption{(a) Simulation results for $S P(S)$ for varying flow rates. (b) Results of the proposed
  model for $P(I)$ with $S_{max}$ increasing from the bottom curve to the top curve.
  The inset shows the ratio of $P(I_{peak})$ to $P(I_{min})$ as a function of $1/S_{max}$.}
  \label{modelPI}
\end{figure}

As detailed previously, the criterion that we use to identify the
clusters in our simulations is that of ``frequent'' collisions.
Due to the inelastic nature of the collisions, the velocities of
these particles become highly correlated\cite{du95} and they can
be idealized as clusters of disks moving with the same velocity.
This velocity correlation within spatial clusters of particles has
been directly observed in granular surface flows\cite{bonamy02}.
Since we observe these clusters to be linear and growing in size
with decreasing flow rates, it is plausible that the changes in
$P(I)$ observed in our simulations is related to a change in
distribution of the size of the clusters of frequently-colliding
disks.   In order to put this conjecture on a firmer footing, we
calculated the impulse distribution from our model using forms of
$P(v)$ and $P(S)$  which provide good fits to our simulation data.
The form of $P(v)$ that we use is a Gaussian with an exponential
tail.  This form is representative of the simulation results for
$P(v)$ with the center of the Gaussian reflecting the average flow
velocity.  The observed cluster size distribution is consistent
with the form $P(S) = \exp(-S/S_{max})/S$ with $S_{max}$
increasing as the flow velocity decreases.  Fig.~\ref{modelPI}a
shows $P(S)$ obtained from our simulations for three different
flow rates.  Note that a change in $\omega_{0}$ will produce a
change in $S_{max}$ for a given flow velocity but will not alter
the shape of $P(S)$ nor the trend of increasing $S_{max}$ with
decreasing flow velocity.

Using these forms of $P(v)$ and $P(S)$, we find that the $P(I)$ obtained
from the simple  model has a form which is remarkably similar to the measured $P(I)$,
as shown in Fig.~\ref{modelPI}b.  The control parameter in the model is $S_{max}$ and
the contribution at small impulses increases with $S_{max}$ as is evident from the
inset to Fig.~\ref{modelPI}b.  The correspondence between the model and our simulations
strongly suggests that the observed changes in $P(I)$ are associated with the growth of
clusters of frequently colliding particles and as $S_{max}$ diverges, the ratio of
$P(I_{min})$ to $P(I_{peak})$ approaches unity.  The tail of the $P(I)$ distribution is
controlled by the shape of $P(v)$ as we have verified within the model.

The picture which seems to be emerging from our simulations is that of increasingly
larger scale spatial heterogeneities developing as the system approaches jamming.
Since the clusters are essentially the same as the ones identified in freely cooling
granular matter, their origin lies in the dissipative nature of the medium.  The
heterogeneities reflect strong velocity correlations of grains and leave a distinctive
signature in the impulse distribution.  Whether the clusters that we have identified
are indeed incipient force chains remains to be verified.  If this connection can
be established, for example through the calculation of stress correlations in the
flowing medium, then our analysis will provide a natural connection between $P(f)$
and force chains.  Moreover, it would a indicate that jamming occurs through the
formation of system spanning clusters of grains whose velocities are strongly correlated.

\paragraph{Acknowledgements}
We thank N. Menon, S. Tewari, N. Easwar and S.R. Nagel for many
helpful discussions. AF, BF and BC acknowledge support from NSF
through  grant No. DMR 0207106, and AF acknowledges support from
the Natural Sciences and Engineering Research Council, Canada.


\begin{thebibliography}{0}

\bibitem{jaeger96} H.M.~Jaeger, S.R.~Nagel and R.P.~Behringer, Rev. Mod. Phys. {\bf 68}, 1259 (1996).
\bibitem{kadanoff99} L.P.~Kadanoff, Rev. Mod. Phys. {\bf 71}, 435 (1999).
\bibitem{liu95} C. H. Liu {\em et al.}, Science {\bf 269}, 513 (1995).
\bibitem{howell99} D.~Howell, R.P.~Behringer and C.~Veje, Phys. Rev. Lett. {\bf 82}, 5241 (1999).
\bibitem{mueth98} D. M.~Mueth, H. M.~Jaeger and S. R.~Nagel, Phys. Rev. E {\bf 57}, 3164 (1998).
\bibitem{ohern01} C.S.~O'Hern, S.A.~Langer, A.J.~Liu and S.R.~Nagel, Phys. Rev. Lett. {\bf 86}, 111 (2001).
\bibitem{mills99} P.~Mills, D.~Loggia and M.~Tixier, Europhys. Lett. {\bf 45}, 733 (1999).
\bibitem{bonamy02} D.~Bonamy {\em et al.}, Phys. Rev. Lett. {\bf 89}, 034301 (2002).
\bibitem{miller96} B.~Miller, C.~O\'Hern and R. P.~Behringer, Phys. Rev. Lett. {\bf 77}, 3110 (1996).
\bibitem{silbert02} L.E. Silbert {\em et al.}, Phys. Rev. E {\bf 65}, 051307 (2002).
\bibitem{longhi02} E.~Longhi, N.~Easwar and N.~Menon, Phys. Rev. Lett. {\bf 89}, 045501 (2002).
\bibitem{mcnamara96} S.~McNamara and W.R.~Young, Phys. Rev. E {\bf 53}, 5089 Part B (1996).
\bibitem{denniston99} C.~Denniston and H.~Li, Phys. Rev. E {\bf 59}, 3289 (1999).
\bibitem{mcnamara94} S.~McNamara and W.R.~Young, Phys. Rev. E {\bf 50}, R28 (1994).
\bibitem{du95} Y.~Du, H.~Li and L.P.~Kadanoff, Phys. Rev. Lett. {\bf 74}, 1268 (1995).

\end{thebibliography}
\end{document}